\documentclass[twocolumn]{aastex6}

\begin{document}

%% LaTeX will automatically break titles if they run longer than
%% one line. However, you may use \\ to force a line break if
%% you desire.

\title{Dynamical Formation of the GW150914 Binary Black hole}

%% Use \author, \affil, plus the \and command to format author and affiliation
%% information.  If done correctly the peer review system will be able to
%% automatically put the author and affiliation information from the manuscript
%% and save the corresponding author the trouble of entering it by hand.
%%
%% The \affil should be used to document primary affiliations and the
%% \altaffil should be used for secondary affiliations, titles, or email.

%% Authors with the same affiliation can be grouped in a single
%% \author and \affil call.

%\normalsize{$^{1}$Center for Interdisciplinary Exploration and Research in Astrophysics (CIERA) }\\
%    \normalsize{and Dept.~of Physics and Astronomy, Northwestern
%      University,}\\\normalsize{2145 Sheridan Rd, Evanston, IL 60208, USA}\\
%\normalsize{$^{2}$School of Physics and Astronomy, University of Birmingham,}\\
%\normalsize{Birmingham, B15 2TT, United Kingdom}\\

\author{Carl L. Rodriguez\altaffilmark{1}, Carl-Johan Haster\altaffilmark{2}, Sourav Chatterjee\altaffilmark{1}, Vicky Kalogera\altaffilmark{1}, and Frederic A. Rasio\altaffilmark{1}}

\altaffiltext{1}{Center for Interdisciplinary Exploration and Research in Astrophysics (CIERA)
and Dept.~of Physics and Astronomy, Northwestern University
2145 Sheridan Rd, Evanston, IL 60208, USA}
\altaffiltext{2}{School of Physics and Astronomy, University of Birmingham,
Birmingham, B15 2TT, United Kingdom}

%% Notice that each of these authors has alternate affiliations, which
%% are identified by the \altaffilmark after each name.  Specify alternate
%% affiliation information with \altaffiltext, with one command per each
%% affiliation.

%\altaffiltext{1}{AAS Journals Data Scientist}
%\altaffiltext{2}{greg.schwarz@aas.org}
%\altaffiltext{3}{AAS Journals Associate Editor-in-Chief}
%\altaffiltext{4}{AAS Director of Publishing}
%\altaffiltext{5}{IOP Senior Publisher for the AAS Journals}

%% Mark off the abstract in the ``abstract'' environment.
\begin{abstract}

We explore the possibility that GW150914, the binary black hole merger recently
detected by Advanced LIGO, was formed by gravitational interactions in the core of a
dense star cluster. Using models of globular clusters with detailed $N$-body dynamics
and stellar evolution, we show that a typical cluster with a mass of
$3\times10^5M_{\odot}$ to $6\times10^5M_{\odot}$ is optimal for forming
GW150914-like binary black holes that will merge in the local universe.  We
identify the most likely dynamical processes for forming GW150914 in such a
cluster, and we show that the detection of GW150914 is consistent with the
masses and merger rates expected for binary black holes from globular clusters.  Our results show that dynamical processes provide a significant and well-understood pathway for forming binary black hole mergers in the local universe.  Understanding the contribution of dynamics to the binary black hole merger problem is a critical step in unlocking the full potential of gravitational-wave astronomy.

\end{abstract}

%% Keywords should appear after the \end{abstract} command.
%% See the online documentation for the full list of available subject
%% keywords and the rules for their use.
\keywords{globular clusters: general --- stars: black holes --- gravitational
waves}

\maketitle

%% From the front matter, we move on to the body of the paper.
%% Sections are demarcated by \section and \subsection, respectively.
%% Observe the use of the LaTeX \label
%% command after the \subsection to give a symbolic KEY to the
%% subsection for cross-referencing in a \ref command.
%% You can use LaTeX's \ref and \label commands to keep track of
%% cross-references to sections, equations, tables, and figures.
%% That way, if you change the order of any elements, LaTeX will
%% automatically renumber them.

%% We recommend that authors also use the natbib \citep
%% and \citet commands to identify citations.  The citations are
%% tied to the reference list via symbolic KEYs. The KEY corresponds
%% to the KEY in the \bibitem in the reference list below.

\section{Introduction}

The first detection of gravitational waves from merging black holes \citep{Abbott2016a} has begun a new era of astronomy.  GW150914 carried with it profound implications for many astrophysical models of black hole (BH) formation throughout cosmic time.  The surprisingly large masses of the individual BH components ($36M_{\odot}$ and $29M_{\odot}$) strongly suggest that the binary BH (BBH) progenitor of GW150914 was formed in a low-metallicity environment \citep[][and references therein]{Abbott2016}. Theoretical studies have shown that dynamical interactions in the dense cores of low-metallicity globular clusters (GCs) are particularly efficient at forming BBH systems similar to GW150914 \citep{Rodriguez2016a,Chatterjee2016}.

Shortly after the formation of a GC, the most massive stars collapse, leaving behind a population of BHs.  Being more massive than the typical star, these BHs rapidly sink into the center of the cluster, where the tremendous spatial density of BHs ($\sim 10^6\, \text{pc}^{-3}$) can produce many encounters between BHs and BH binaries.
These encounters can create new binaries in one of two ways: either three single BHs pass
sufficiently close to one another for two BHs to become bound (with the third BH receiving a
velocity boost to conserve energy), or an exchange interaction occurs in which a BH
 inserts itself into a preexisting binary, ejecting one of its
components. The newly-forged binaries remain in the GC core, where they
are continuously modified, disrupted, and recreated by encounters with other binary and single BHs.
This dynamical mosh pit continues, forming more and more
strongly bound binaries \citep{Heggie1975} until a BBH receives a sufficient velocity kick to
eject it completely from the cluster. In effect, the extreme density of BHs
in a GC core creates a veritable gravitational-wave factory,
continuously forming and ejecting BBHs into the field to await their fate as Advanced LIGO sources.

In this letter, we explore the dynamical formation channel for BBHs in the context of
GW150914.
%questionsthe possibility that GW150914 was forged in the core of a globular cluster (GC) by three- and four-body gravitational encounters.
%Here we address three very specific, well-constrained questions:
In Section \ref{sec:gc}, we explore what
type of present-day GC could have formed the GW150914 BBH.  In
Section \ref{sec:formation}, we show what
type of interactions in a GC are responsible for the dynamical formation of
GW150914-like binaries.  Finally, in Section \ref{sec:detection}, we show that the
GW150914 detection is consistent with
the masses and redshifts of BBHs from GCs that we expect Advanced LIGO to
detect.

To answer these questions, we use 48 detailed star-by-star GC models, developed in
\cite{Rodriguez2016a}, which incorporate all the relevant dynamical physics and
advanced algorithms for stellar evolution, equivalent to the most recent
studies of BBH mergers from galactic fields \citep{Dominik2013}.  These models
span large ranges in initial masses ($N=2\times10^5,~5\times10^5,~1\times10^6,$ and $2\times
10^6$ initial particles, corresponding to present-day masses of $\sim
5\times10^4M_{\odot}$, $ 1\times10^5M_{\odot}$, $3\times10^5M_{\odot}$,
and $6\times10^5M_{\odot}$, respectively), stellar metallicities
($0.25Z_{\odot},~0.05Z_{\odot},$ and $0.01Z_{\odot}$), and initial virial radii
(1 pc and 2 pc). See Appendix \ref{sec:app1} for more details.

\section{Forming Heavy BBHs in GCs}
\label{sec:gc}

We extract from our 48 models all the binaries that appear similar to GW150914.   We start by looking at any BBH whose source-frame component and chirp masses fall within the 90\% credible regions for GW150914 \cite[$m_1 = 35.7^{+5.4}_{-3.8}M_{\odot}$, $m_2 = 29.1^{+3.8}_{-4.4}M_{\odot}$, and $\mathcal{M}_c = 27.9^{+2.1}_{-1.7}M_{\odot}$, from][]{TheLIGOScientificCollaboration2016}.  This corresponds to a total of 262 BBHs from 40 of the 48 GC models, 259 of which merge outside the cluster.  We assume all GCs formed $\sim 12$ Gyr ago (at $z\simeq 3.5$, consistent with GCs in the Milky Way, although other galaxies, such as the Large and Small Magellanic Clouds, have significantly younger GC populations).  Of the 8 GC models that do not contribute BBHs with masses like GW150914, 4 have disrupted before 12 Gyr and are exlcluded from our analysis, and the remaining 4 have low initial $N$ and lower number of initial BHs.  The remaining 40 GC models contribute roughly equal numbers of GW150914-like BBHs (when normalized to the number of initial stars in each model).  Our models show a strong dependence on metallicity, with the $Z = 0.05Z_{\odot}~\rm{and}~0.01Z_{\odot}$ models contributing nearly 3 and 5 times as many BBHs as the $Z=0.25Z_{\odot}$ models, respectively.

We then define a true GW150914 progenitor to be the subset of these 262 binaries that merge between 7 and 13 Gyr after GC formation, corresponding to mergers that occur in the local universe ($z < 0.5$).  We find 14 such systems across our 48 models, all of which were ejected from the cluster prior to merger.  Of these 14, we find that 10 originate from 5 GC models with similar initial conditions, corresponding to GCs with lower metallicities ($0.05Z_{\odot}$ and $ 0.01Z_{\odot}$, typical for the low-metallicity clusters in most galaxies), large masses ($N = 1\times10^6~\rm{and}~2\times10^6$ initial
particles, corresponding to final masses of $3\times10^5M_{\odot}~\rm{to}~6\times10^5M_{\odot}$ today), and typical virial radii ($R_v = 2$ pc).  That these binaries (and the majority of all 262 GW150914-like BBHs) form from
low metallicity and massive clusters is unsurprising: lower metallicities yield
less effective stellar winds \citep{Vink2011}, reducing the amount of mass that is
lost before a massive star collapses, and producing ``heavy'' BHs like the observed components of GW150914 \citep{Belczynski2010,Mapelli2013,Spera2015a}.  Furthermore, massive clusters produce
a larger number of BHs, which enhances the dynamical production of BBHs.

\begin{figure}[tbh!]
\centering
\includegraphics[trim=2cm 0cm 2cm 0cm,scale=0.7]{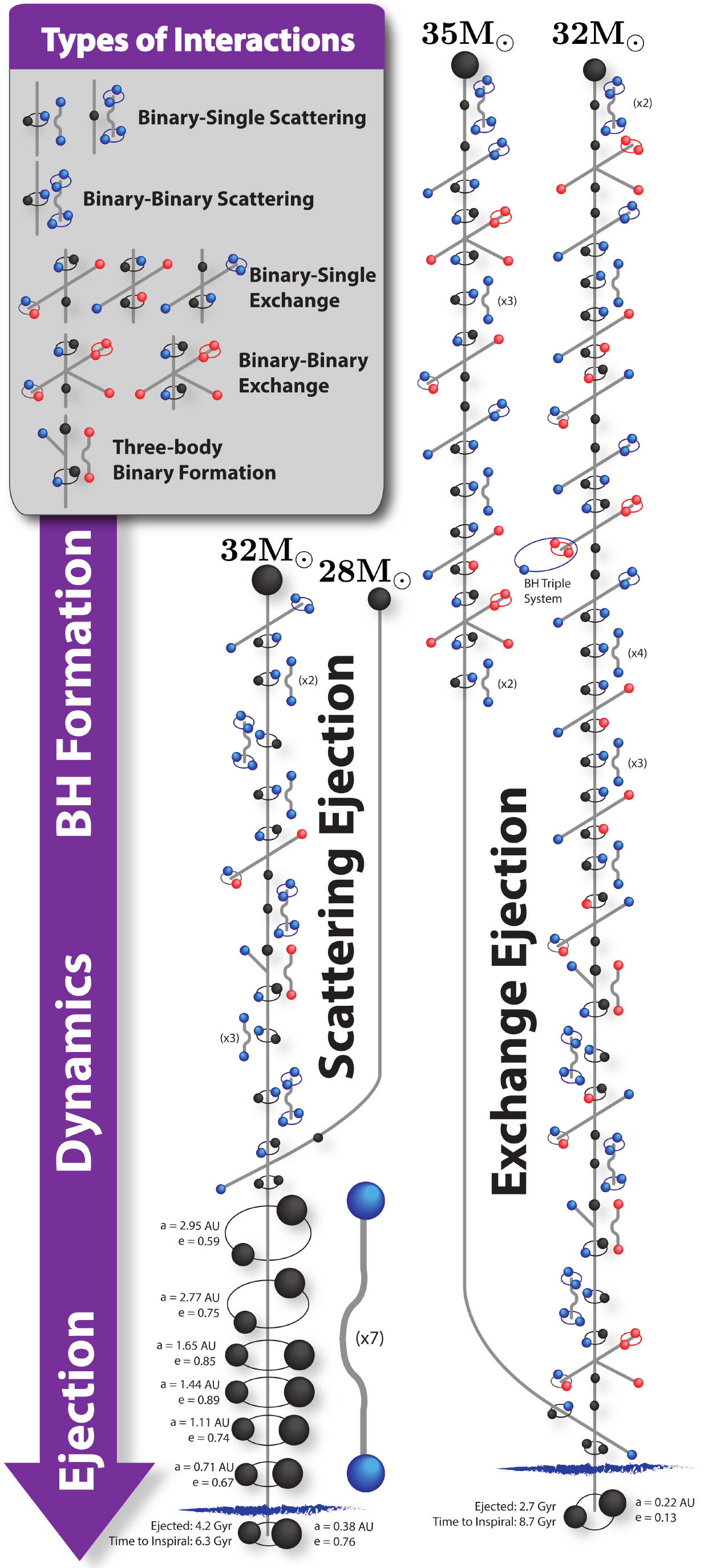}
\caption{Interaction diagram showing the formation history for two GW150914 progenitors in a single GC model.  From top to bottom, the history of each individual BH that will eventually comprise a GW150914-like binary is illustrated, including all binary interactions.  The legend shows the various types of gravitational encounters included in our GC models (with the exception of two-body relaxation).  In each interaction, the black sphere represents the GW150914 progenitor BH, while the blue and red spheres represent other BHs (and stars) in the cluster core.}
\label{fig:theLongthing}
\end{figure}

%\section*{Detecting Binary BHs from GCs
The preference for clusters with larger virial radius (2 pc versus the more compact 1 pc clusters) arises from the need for
long inspiral times.  Binaries with total masses of $\sim 60M_{\odot}$ are more massive than the average stellar or BH mass in the cluster, and are typically ejected within the first few Gyrs of a cluster's
evolution.  However, since GW150914 merged $\sim 1.3$ Gyr ago ($\gtrsim
10$ Gyr after the formation of the old GCs considered here), it must have been ejected from a
cluster environment with a sufficiently wide separation to ensure a delay
time of $\sim 10$ Gyr before merger.  It is a well-known result \citep{Zwart1999,Moody2009}
that, despite the chaotic nature of dynamical formation, it is the global cluster properties that primarily determine the semi-major axis of binaries at ejection.  In \cite{Rodriguez2016a}, we showed that this relationship can be expressed as

\begin{equation}
\frac{R_v}{M_{GC}} \sim \frac{a}{\mu_{\text{bin}}}
\label{eqn:equiv}
\end{equation}

\noindent where $M_{GC}$ and $R_v$ are the mass and virial radius of the
cluster, and $a$ and $\mu_{\text{bin}}$ are the semi-major axis and reduced mass
of the binary.  Equation (\ref{eqn:equiv})  shows that, for a given binary mass, more massive clusters must have large virial radii to produce binaries with large semi-major axes\footnote{{Note that the proportionality constant in \ref{eqn:equiv} can vary from $\sim 10$ to $\sim 100$ (in solar units) within a fixed cluster.  See \cite{Rodriguez2016a}, Figure 2 and Equations 6-10.}}.  This result holds true in our models: the massive GCs with $R_v = 1$ pc produce $\sim 60M_{\odot}$ BBHs at a rate similar to GCs with $R_v = 2$ pc; however, the majority of binaries from those compact clusters are ejected
within the first Gyr of the cluster evolution and merge $\lesssim 1$ Gyr later.
For the binaries to merge in the local universe, they were most likely ejected
from a massive cluster with a virial radius $\sim 2$ pc.   We conclude that,
were it formed dynamically, the progenitor of GW150914 most likely originated in
a low-metallicity GC with a present-day mass between $3\times 10^5M_{\odot}$ and
$6\times 10^5 M_{\odot}$ and an  initial virial radius of 2 pc, typical of young
clusters in the local universe \citep[e.g.,][]{Scheepmaker2007}.

\section{Dynamical formation of GW150914}
\label{sec:formation}

In addition to the statistics of the ejected BBHs, our GC models allow us to describe the specific dynamical interactions that created a potential GW150914 BBH.
None of our 14 GW150914 progenitors are formed from
primordial stellar binaries that become BBHs, and only 12 of all 262 binaries with GW150914-like
masses are formed directly from a primordial binary.  Instead, all but one of the 14 progenitors
 were created during a strong gravitational exchange encounter involving
 either one binary and one single BH (in 11 cases) or two BBHs (in 2
 cases).  Only one binary was created by an interaction involving three single
 BHs \citep[a ``three-body binary'' formation,][]{Binney2011}.  This result is
 surprising, given that three-body binary formation is expected to be the
 dominant mechanism for creating new BBHs in the cores of GCs
 \citep{Morscher2015}.  However, these three-body binaries are not necessarily the same binaries
 that will become future gravitational-wave sources. In
 order to be ejected from the cluster, this first generation of binaries must
 undergo several scattering encounters to pump up their gravitational binding
 energies--encounters that offer many opportunities to form new binaries by
 exchanging components.  These gravitational encounters erase the original state of the first generation of BBHs, and become the primary mechanism for producing BBH mergers from GCs.
 
 \begin{figure}[bh!]
\centering
\includegraphics[scale=0.6]{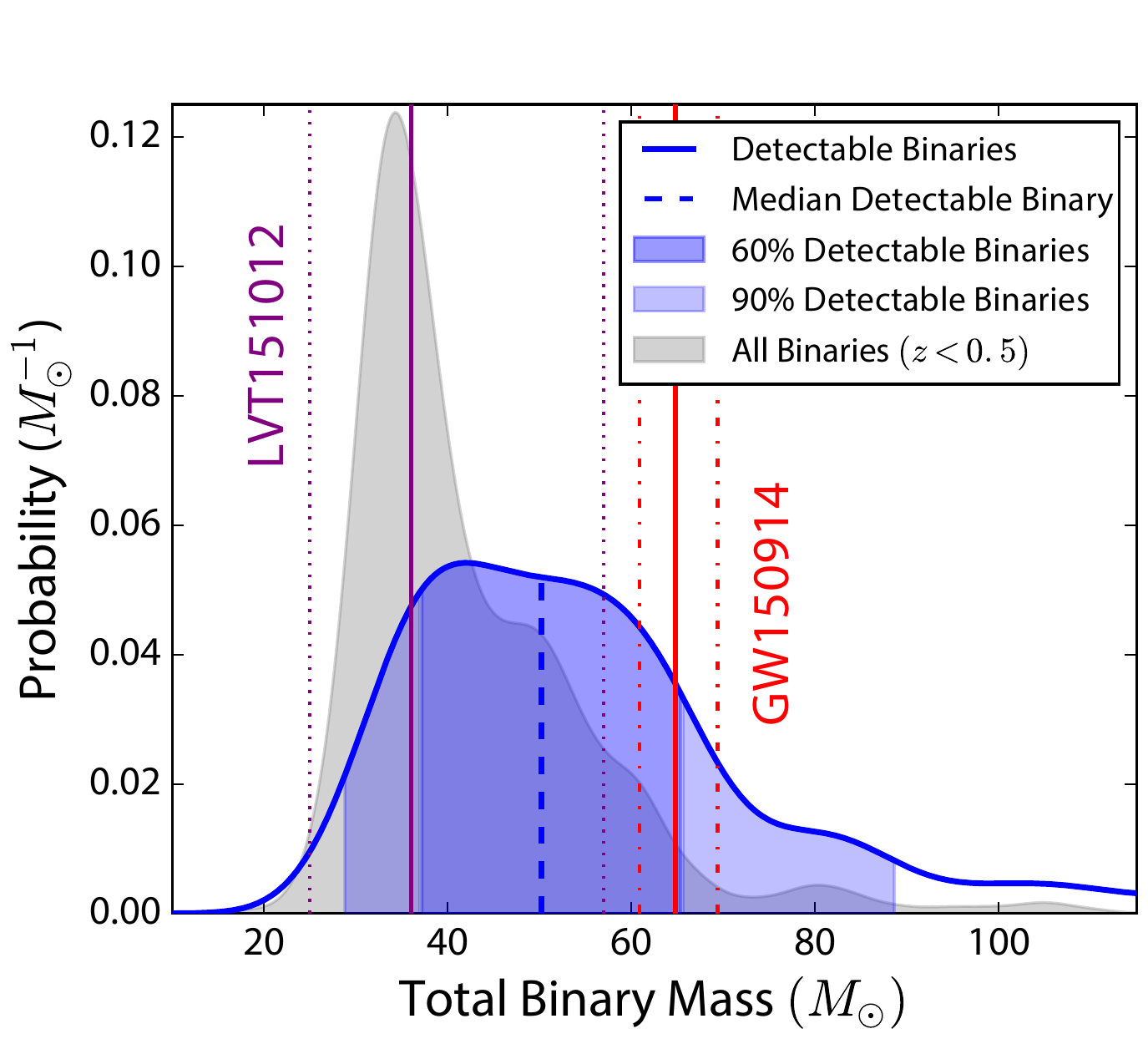}
\caption{The distribution of BBH total masses from GCs.  In gray, we show the
distribution of all mergers that occur at $z<0.5$ (for GCs that form at
$z\simeq3.5$), while in blue we show the distribution of sources detectable with
Advanced LIGO during its first observing run.  The median and 90\% credible
regions for the total mass of GW150914 are shown in red
\citep{TheLIGOScientificCollaboration2016}.  We also show the gravitational-wave
trigger, LVT151012, in purple \cite[where we have computed the median and
credible regions by adding the component mass median and 90\% credible interval
boundaries from][]{TheLIGOScientificCollaboration2016a}.  Note that, while LVT151012 is below the threshold to be considered a detection, there exists a $\gtrsim 84\%$ chance that the signal was of astrophysical origin \citep{Abbott2016b}.}
\label{fig:detect}
\end{figure}

 \begin{table*}[tbh!]
\begin{tabular*}{\textwidth}{@{\extracolsep{\fill}} l|ccc}
\hline\hline
\textbf{Mergers} & \textbf{Pessimistic} & \textbf{Realistic} & \textbf{Optimistic} \\\hline
\hline

O1 (Detections / 16 Days)&0.05&0.2&0.7\\
O1 (Detections / 50 Days)&0.2&0.5&2\\
O2 (Detections / Year)&4&15&60\\
Design Sensitivity (Detections / Year)&30&100&400\\\hline\hline
Merger Rate at $z\sim 0.1$ (Mergers / Gpc$^{3}$ / Year) &2&5&20 \\
Merger Rate at $z\sim 1$ (Mergers / Gpc$^{3}$ / Year) &4&10&40 \\\hline\hline
\end{tabular*}
\caption{The expected merger rate for all BBHs from GCs.  We show the theoretical detection rate for the first observing run of Advanced LIGO (O1) over a 16 day period (consistent with the GW150914 detection) and over a $\sim$50 day period (the length of O1, assuming a $\sim 4$ month duration \citep{Abbott2016c} with a double-coincident runtime fraction of $16/39$ \citep{Abbott2016b}).  We also show the detection rate given the projected sensitivity for Advanced LIGO's second observing run (O2, with a proposed length of 6 months \citep{Abbott2016c}, for which we use the mid-sensitivity curve from \cite{O2sensitivity-T1200307}), and the final design sensitivity from \cite{AdvSensitivity-T0900288-v3}.  Both projected rates assume a year of double-coincident data from both LIGO detectors.  For reference, we show the total merger rate density from \cite{Rodriguez2016a} at $z \sim 0.1$ (the observed redshift of GW150914) and at $z\sim 1$.  The optimistic and pessimistic rates are computed assuming the $\pm 1 \sigma$ uncertainties on the spatial density of GCs in the universe from \cite{Rodriguez2015a}, and considering all GCs to have initial virial radii of 1 pc or 2 pc, respectively.  The realistic rate assumes the mean spatial density of GCs, and an even mix of 1 pc and 2 pc clusters.}
\label{tab:models}
\end{table*}

In addition to their formation, all 14 GW150914 progenitors were ejected from their host clusters after a strong interaction.  This is consistent with
 \cite{Rodriguez2016a}, which found that $81\%$ of BBHs ejected from a GC are
 ejected following a binary-single encounter, and $13\%$ following a
 binary-binary encounter.  We find that 9 of the 14 binaries were ejected from the cluster
 following an exchange encounter, in which preexisting binary exchanged components
 and was ejected from the cluster before it could interact with other BHs
 (although the binary that was exchanged into may have undergone many encounters
 before ejection).  The remaining 5 binaries were retained in the cluster after formation, and continued to interact with other BHs in the core until one such encounter ejected the BBH from the cluster.  We show two examples of GW150914 formation histories in Figure \ref{fig:theLongthing}, one in which the binary is ejected following an exchange encounter, and one in which the binary is formed and ejected after repeated scattering encounters.
 
Binaries that are
 ejected following an exchange interaction tend to be ejected from the cluster early,
 when there are still many $\sim 30M_{\odot}$ BHs in the cluster.  Conversely, systems that are ejected by repeated
 scatterings interactions tend to be ejected several Gyr later, when there are fewer BHs of comparable
 mass to produce an energetic exchange interaction.  The distribution of
 inspiral times from both processes is nearly identical.  But since we are only
 interested in GW150914-like binaries that merge in the local universe, we find that binaries ejected following an exchange are ejected early with longer inspiral times, while binaries ejected following several scattering interactions are ejected later with shorter inspiral times.

 As Figure \ref{fig:theLongthing} makes clear, the dynamical history of any particular system is quite complex.  But the interactions ensure that the orbital properties of dynamically-formed BBHs are a function only of well-understood gravitational processes, completely free of any dependence on the initial conditions of the BBH population.  This eliminates many of the uncertainties associated with the modeling of isolated binary stellar evolution in galactic fields.  The dynamical formation channel is largely independent of the many unconstrained parameters of binary evolution (e.g.~the outcome of common envelope evolution) that can cause estimates of the BBH merger rate from the field to vary by several orders of magnitude \citep{Rodriguez2016a}.

\section{Detection Rate}
\label{sec:detection}

With this understanding of the dynamical formation scenario, it is only natural
to ask: what masses of dynamically-formed BBHs are most likely to be detected by
Advanced LIGO?  The answer depends on two factors: the underlying distribution
of BBH mergers in mass and redshift, and the sensitivity of the LIGO detector to BBH
mergers with specific masses at a given redshift.  In Figure \ref{fig:detect}, we show the
distribution of BBH mergers from all our models, with the BBHs drawn randomly
from specific GC models proportionally to the observed mass distribution of GCs
\citep[with clusters closer to the peak of the GC mass function contributing more BBH mergers to our effective sample, see ][and Appendix \ref{sec:app2}]{Harris2014,Rodriguez2016a}.  {Although there exist many mergers in the local universe ($z<0.5$) with total masses from $20M_{\odot}$ to $120_{\odot}$, the majority of mergers occuring in the present day lie in the peak between $30M_{\odot}$ and $40M_{\odot}$.  This is consistent with \cite{Morscher2015,Rodriguez2016a}, which found that GCs process through their most massive BHs early, leaving behind the less-massive systems to form binaries and merge in the local universe.  The peak at $\sim 35M_{\odot}$ is primarily dominated by contributions from the $Z=0.25Z_{\odot}$ models, while the tail extending to high masses is primarily from low-metallicity ($Z=0.05Z_{\odot},~0.01Z_{\odot}$) clusters.  As with GW150914, our models show that mergers more massive that $40M_{\odot}$ at low redshifts are most likely to have been formed in massive, low-metallicity clusters.}

To translate this into a distribution and rate of detectable sources, we combine the total distribution of BBH mergers with the publicly-available
Advanced LIGO sensitivity spectrum representative for the GW150914 observation
\citep[][]{O1sensitivity-G1501223} and a  and compute the distribution of detectable BBHs from GCs.  We find that the median total mass of a BBH detectable during the 16 days of Advanced LIGO's first observing run (O1) is $ 50M_{\odot}$, with 60\% of sources having total masses from $37M_{\odot}$ to $66M_{\odot}$ (enclosing the $65M_{\odot}$ total mass of GW150914), and 90\% of sources having masses from $29M_{\odot}$ to $89M_{\odot}$.  In Table 1, we integrate the mass distribution over all redshifts, and list the detection rate of BBH mergers from GCs for different current and planned observing runs of Advanced LIGO.
We find that, during the first 16 days of O1, Advanced LIGO could have detected anywhere from 0.05 to 0.7 BBH mergers from GCs.  Based on these results, we conclude that GW150914 is consistent with dynamical formation in a GC.

With only a single detection, and significant uncertainties on the BBH
 merger rate from isolated binary stellar evolution, it cannot be definitively said
 which of the many proposed formation channels produced GW150914.  However, both GW150914 and the results presented here indicate that Advanced LIGO may detect many more BBH mergers in the near future \citep{Abbott2016b}.  Once Advanced LIGO has produced a catalog of
 BBH merger candidates with different masses and spins at different redshifts, we
 will begin to constrain many of the existing BBH population models, yielding tremendous information about BH formation and dynamics across cosmic
 time.

\acknowledgments{
We thank Ilya Mandel for carefully reviewing this manuscript, Chris Pankow for useful discussions, and the referee for their useful suggestions.  This work was supported by NSF Grant AST-1312945 and NASA Grant NNX14AP92G.
}

\bibliography{biblio.bib}
\bibliographystyle{aasjournal}

\appendix
\section{Monte Carlo Simulations of Globular Clusters}
\label{sec:app1}

Throughout this study, we use a series of 48 GC models, first created in
\cite{Rodriguez2016a}, to explore the dynamical formation pathways of GW150914.
These models were created with our Cluster Monte Carlo (CMC) code, a H\'enon-style
Monte Carlo approach to stellar dynamics \citep{Henon1971,Henon1975}.  By
assuming spherical symmetry, a large number of particles, and dynamics
primarily driven by two-body relaxation, CMC can model GCs with significantly
more single and binary objects than a direct $N$-body integration.  Recent work
\citep{Rodriguez2016} has shown that CMC can model GCs with $N\sim 10^6$
particles with similar accuracy to state-of-the-art direct integrators,
reproducing both the global cluster parameters and the BBH properties in a
fraction of the time, allowing us to fully explore the parameter space of dense star clusters.

\begin{figure}[htp]
\centering
\includegraphics[trim=4cm 0cm 0cm 0cm,scale=0.55]{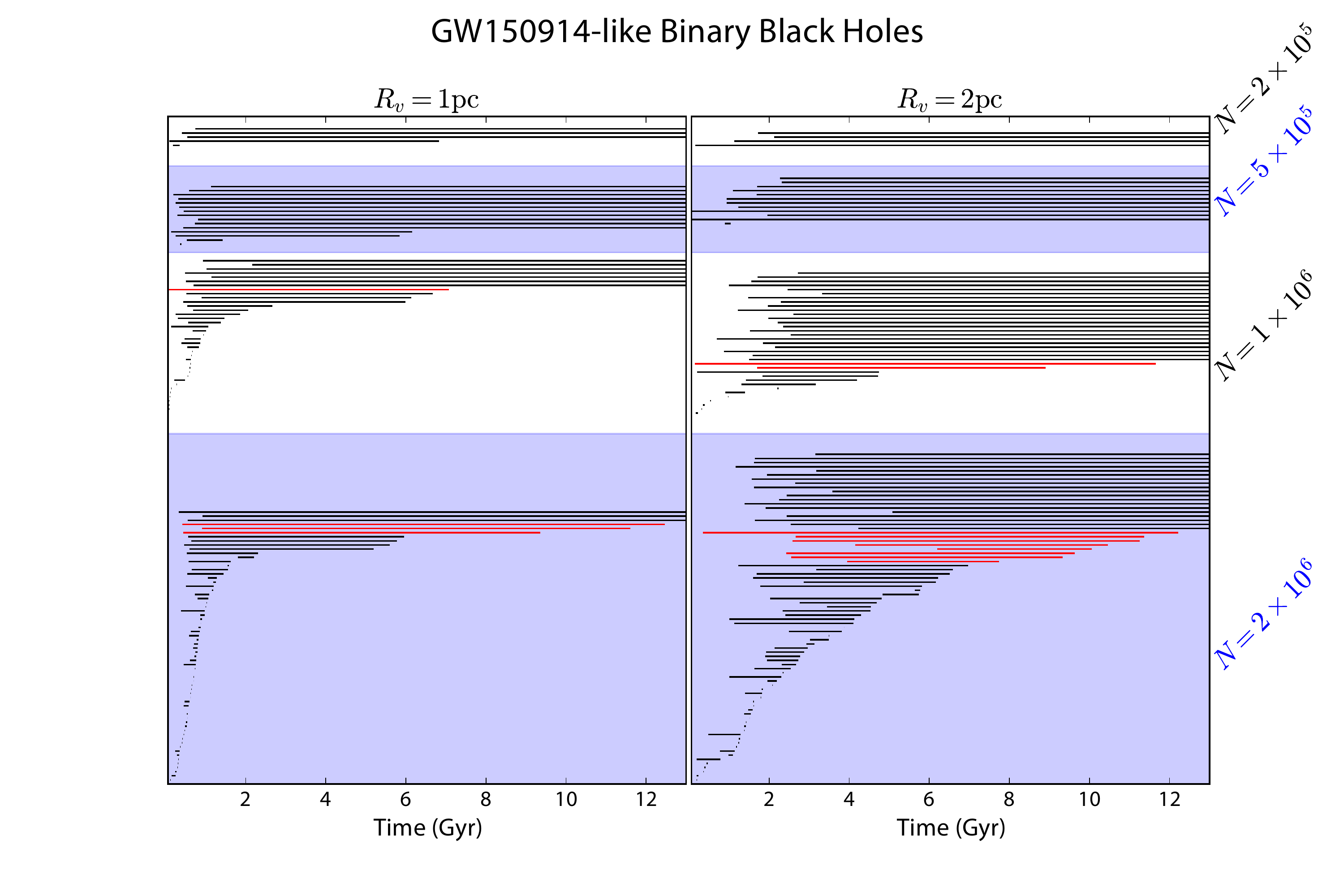}
\caption{Graphical representation of all 262 BBHs with masses similar to GW150914 from the 48 GC models, sorted in order of increasing merger times.  {We separate the models according to their initial $N$ and virial radii, such that each rectangle represents the BBHs from 6 models.}  The left end of each line indicates when each binary was ejected from the cluster and the right end indicates the merger time.  Those binaries that merge in the local universe ($z < 0.5$, which we consider GW150914 progenitors) are indicated in red.  {As we consider more massive and more compact clusters, a larger number of BBHs will merge within a Hubble time, consistent with previous results \cite[][Figure 1]{Rodriguez2016a}.}}
\label{fig:thething}
\end{figure}

In addition to two-body relaxation, various improvements to CMC have
incorporated {much of the necessary} physics to correctly treat the
 BBH dynamical formation problem.  In particular, we consider:

\begin{itemize}
\item \textbf{three-body binary formation}, using a probabilistic prescription
for three BHs that pass within a certain radius \citep{Morscher2012},
\item \textbf{strong three-body and four-body gravitational scattering
encounters}, computed by direct summation of Newtonian gravity with the
\texttt{Fewbody} integrator \citep{Fregeau2004}, and
\item \textbf{single and binary stellar evolution}, using the BSE package for
rapid stellar evolution \citep{Hurley2000,Hurley2002}.
\end{itemize}

\noindent We have enhanced our stellar
evolution prescriptions with new physics describing the distribution of remnant
masses for BHs formed
from core-collapsed massive stars, metallicity and temperature-dependent stellar winds, and the natal
kicks of BHs.  See \cite{Rodriguez2016a} for details.  These prescriptions are identical to those employed in the most recent estimates of BBH merger rates
from galactic fields \citep{Dominik2013}, allowing us to
directly compare our dynamical results to other channels for BBH formation.  Note that we do not consider relativistic effects, such as post-Newtonian corrections to close dynamical encounters \citep{Samsing2014} or long-term secular effects, such as Lidov-Kozai and gravitational wave emission in BH triple systems \citep{Antonini2014,Antonini2015}.  Although these only contribute at the $\sim 1\%$, they should still be considered for any dynamically-complete treatment.  We also assume that GCs do not contain intermediate-mass BHs which can influence the BBH formation and merger rate \cite[e.g.,][]{Macleod2016}.

Our 48 cluster models are generated with a different number of initial particles, virial radii, and
metallicities, in order to explore the full range of massive GCs
observed in the Milky Way and other galaxies.  We consider a grid with three different stellar
metallicities ($Z = 0.25Z_{\odot},0.05Z_{\odot},$ and $0.01Z_{\odot}$), two different
initial virial radii ($R_v = 1,2$ pc) and four different initial particle
numbers ($N=2\times10^5,5\times10^5,1\times10^6,2\times10^6$).  Each set of initials conditions was run twice, with a different initial state of particle positions and velocities, for a total of 48 models.  The details of these models can be found in \cite{Rodriguez2016a}.

Finally, we list the properties of the 14 GW150914 progenitors we consider in this study.  In Table 2, we list the masses and orbital properties for each of the binaries, while in Table 3, we list the initial conditions of the clusters that created each BBH, and information about the dynamics of the binary inside the cluster.  In Figure \ref{fig:thething}, we show all 262 BBHs from our models with masses similar to GW150914, highlighting in red those 14 BBHs that merge in the local universe.

\begin{table*}[t]
\begin{tabular*}{\textwidth}{@{\extracolsep{\fill}} llllllllllll}

%& \textbf{Initial Conditions} &&& \textbf{Properties (12 Gyr)} \\\hline
\# & $m_1$ $(M_{\odot})$ & $m_2$ $(M_{\odot})$  & $a$ (AU)& $e$ & $P_{\text{orb}}$ (days)& $T_{\text{ejec}}$ (Gyr) & $T_{\text{insp}}$ (Gyr) & Redshift ($z$) \\ \hline\hline

1 &38.5 & 26.4 & 0.30 & 0.53&  7.45  & 0.42 & 12.04 & --\\
2 &37.6 & 25.4 & 0.25 & 0.44&  5.75  & 0.44 & 8.92 & 0.21 \\
3 &32.0 & 31.6 & 0.25 & 0.37&  5.72  & 0.91 & 10.70& 0.03 \\
4 &35.3 & 30.5 & 0.40 & 0.74&  11.39 & 0.08 & 7.00 & 0.46  \\
5 &35.5 & 25.8 & 0.21 & 0.13&  4.49  & 0.15 & 11.51& 0.02 \\
6 &37.6 & 30.1 & 0.77 & 0.89&  29.99 & 1.70 & 7.20 & 0.26  \\
7 &36.3 & 28.2 & 0.57 & 0.82&  19.57 & 0.35 & 11.87 & --  \\
8 &32.9 & 32.3 & 0.40 & 0.75&  11.44 & 2.56 & 6.78 & 0.21  \\
9 &32.6 & 32.1 & 0.30 & 0.58&  7.46  & 2.59 & 8.67 & 0.05  \\
10&32.7 & 28.6 & 0.37 & 0.78&  10.50 & 6.20 & 3.86 & 0.15  \\
11&35.1 & 32.5 & 0.22 & 0.13&  4.58  & 2.66 & 8.71 & 0.05 \\
12&33.4 & 27.5 & 1.32 & 0.96&  70.97 & 3.95 & 3.79 & 0.38  \\
13&32.2 & 28.2 & 0.38 & 0.76&  11.01 & 4.15 & 6.30 & 0.12  \\
14&32.9 & 31.3 & 0.25 & 0.50&  5.70  & 2.43 & 7.20  & 0.19 \\

\end{tabular*}
\caption{The 14 GW150914 progenitors identified from 48 GC models.  We list the masses, orbital properties, and inspiral times for each binary at the moment it is ejected from the cluster.  We also show the redshift of each merger, assuming all GCs to be exactly 12 Gyr old.}
\label{tab:models}
\end{table*}

\begin{table*}[t]
\begin{tabular*}{\textwidth}{@{\extracolsep{\fill}} llllllllllll}

%& \textbf{Initial Conditions} &&& \textbf{Properties (12 Gyr)} \\\hline
\# & Metallicity $(Z_{\odot})$ & Particle Number & $R_v$ (pc) &$N^{\text{prev}}_{\text{part}}$ & $N_{BS}$ & $N_{BB}$ & Ejected By\\ \hline\hline
1  & 0.05 & $2\times10^6$ & 1& 0,5&1&0 & BS (Exchange) \\
2  & 0.05 & $2\times10^6$ & 1& 2, 0 & 1 & 0 & BS (Exchange) \\
3  & 0.05 & $2\times10^6$ & 1& 6, 1 & 12 & 3 & BS\\
4 &  0.25 & $1\times10^6$ & 1 & 4, 1 & 1 & 0 & BS (Exchange)\\
5 &  0.05 & $1\times10^6$ & 2 & 0, 1 & 1 & 0 & BS (Exchange) \\
6 &  0.01 & $1\times10^6$ & 2  & 1, 11 & 0 & 1 & BB (Exchange)\\
7 &  0.01 & $2\times10^6$ & 2 & 1, 0 & 1 & 0 & BS (Exchange)\\
8 &  0.01 & $2\times10^6$ & 2  & 6, 1 & 3 & 1 & BS\\
9 &  0.01 & $2\times10^6$ & 2  & 2, 5 & 0 & 1 & BB (Exchange)\\
10 &  0.01 & $2\times10^6$ & 2  & 14, 1 & 1 & 2 & BB\\
11 &  0.01 & $2\times10^6$ & 2  & 8, 4 & 1 & 0& BS (Exchange)\\
12 &  0.01 & $2\times10^6$ & 2  & 0, 12 & 3 & 0& BS \\
13 &  0.01 & $2\times10^6$ & 2  & 2, 0 & 8 & 0& BS \\
14 &  0.05 & $2\times10^6$ & 2 & 11, 5 & 1 & 0& BS (Exchange)\\

\end{tabular*}
\caption{The properties of the clusters that formed each GW150914 progenitor, and the dynamical interactions that formed each binary.  For each, we show the number of previous binary partners each component had before finding its eventual GW150914 companion.  We also show the number of binary-single and binary-binary encounters each binary underwent (including the interaction that created the binary).  Finally, we list the type of encounter that ejected the binary from the cluster, and whether or not that encounter involved an exchange.}
\label{tab:models}
\end{table*}

\section{Gravitational Wave Detectability}
\label{sec:app2}

Computing the detectability of sources requires an understanding of both the sensitivity of the gravitational-wave detector and the underlying mass distribution of BBHs.  Following \cite{Belczynski2014,Rodriguez2015a}, we write the detection rate per unit chirp mass as

\begin{equation}
\mathcal{R}(\mathcal{M}_c) = \int^{\infty}_{0}\mathcal{R}(\mathcal{M}_c,z)f_d(\mathcal{M}_c,z)\left(\frac{dV_c}{dz}\right) \left(\frac{dt_s}{dt_o}\right) dz
\label{eqn:int}
\end{equation}

\noindent where
\begin{itemize}
\item $\mathcal{R}(\mathcal{M}_c,z)$ is the rate of binary mergers at redshift $z$ and chirp mass $\mathcal{M}_c$ from GCs in units of Mpc$^{-3}$ $M_{\odot}^{-1}$ yr$^{-1}$,
\item $f_d(\mathcal{M}_c,z)$ is the fraction of detectable sources at a given chirp mass and redshift,
\item $\left(\frac{dV_c}{dz}\right)$ is the comoving volume at a given redshift, assuming a flat $\Lambda$CDM cosmology with $\Omega_M = 0.306$ and $H_0 = 67.9\text{km}\,\text{s}^{-1}\,\text{Mpc}^{-1}$ \citep{PlanckCollaboration2015}, and
\item $\left(\frac{dt_s}{dt_o}\right)\equiv \frac{1}{1+z}$ is the time dilation between a clock measuring the merger rate at the source and a clock on Earth.
\end{itemize}

To compute the merger rate, $\mathcal{R}(\mathcal{M}_c,z)$, we follow a similar procedure to \cite{Rodriguez2015a,Rodriguez2016a}.  We first assume that all GCs are exactly 12 Gyr old.  The rate is then broken apart into three components:

\begin{equation}
\mathcal{R}(\mathcal{M}_c,z) \equiv \rho_{GC} \left<N_{\rm{insp}}\right> P(\mathcal{M}_c,z)~
\label{eqn:rateThang}
\end{equation}

\noindent where $\rho_{GC}$ is the spatial density of GCs in the universe, which we assume to be $0.77~\rm{Mpc}^{-3}$, with an optimistic value of $2.31~\rm{Mpc}^{-3}$ and a pessimistic value of $0.32~\rm{Mpc}^{-3}$ based on recent measurements of galaxy luminosity functions \citep{Kelvin2014} and the correlation between galaxy luminosity and GC number \citep{Harris2013}.  See \cite{Rodriguez2015a}, supplemental materials, for details of the computation.  We assume $0.77~\rm{Mpc}^{-3}$ to be the standard value, but consider the optimistic and pessimistic values better constrain the uncertainty of our estimate.

{$\left<N_{\rm{insp}}\right>$ is the mean number of BBH mergers produced by a GC over 12 Gyr.  This is found by creating a functional fit of the number of BBH mergers a cluster with a given present-day mass produces over its 12 Gyr lifetime.  In \cite{Rodriguez2016a}, we found a relationship of the form $N_{\rm{insp}}(M_{GC})=N_{\rm{BBH}}(M_{GC})\times f_{\rm{insp}}(M_{GC})$ to be satisfactory, where $N_{\rm{BBH}}(M_{GC})$ is the linear relationship between the cluster mass and the total number of BBHs it produces and $f_{\rm{insp}}(M_{GC})$ is the fraction of BBHs that will merge within 12 Gyr (which we fit to an error function\footnote{Chosen as a simple example of a cumulative distribution function, since the fraction of binaries that will merge within 12 Gyr goes from 0 to 1 as one considers clusters with larger masses.  See \cite{Rodriguez2016a}, and Figure \ref{fig:thething}}).  We then determine $\left<N_{\rm{insp}}\right>$ by integrating the functional form of $N_{\rm{insp}}(M_{GC})$ over the GC mass function, which we take to be a log-normal distribution with mean $\log_{10}(M_0)=5.54$ and width $\sigma_M = 0.52$, based on recently-observed GC luminosity functions in brightest-cluster galaxies \citep{Harris2014} and assuming a mass-to-light ratio of 2 for old stellar systems with minimal dark matter \citep{Bell2003}.  As in \cite{Rodriguez2016a}, we compute three different values of $\left<N_{\rm{insp}}\right>$: the standard value (260 inspirals per 12 Gyr) determined by fitting $N_{\rm{insp}}(M_{GC})$ to all 48 GC models, an optimistic value (347 inspirals per 12 Gyr), found when fitting only to clusters with $R_v = 1~\rm{pc}$, and a pessimistic value (192 inspirals per 12 Gyr), found when considering only clusters with $R_v = 2~\rm{pc}$.  As with the values of $\rho_{GC}$, these estimates will be used to understand the uncertainties associated with our model assumptions.

We determine $P(\mathcal{M}_c,z)$, the probability distribution of mergers
in chirp mass and redshift,  by computing a 2D kernel density estimate (KDE) in
$\mathcal{M}_c$ and binary merger time.  The KDE was generated from our 48 GC
models by sampling binaries at random from each model, appropriately reweighted
to reproduce the mass distribution of GCs observed in the local universe.  This
is accomplished by selecting more binaries from models with larger weights,
where the cluster weights are determined by evenly binning the log-normal GC
mass function so that the average mass of models with the same initial particle
number lies in the midpoint of each bin, then assigning to each model the
integral of the GC mass function over that bin (normalized to the largest
weight).  We compute the weights separately for high- and low-metallicity
clusters, as our high-metallicity models with $N=2\times10^5$ disrupt before 12
Gyr and are excluded from the present analysis.  In practice, this weighting
scheme selects $100\%$ of the binaries from our $N=2\times10^6$ models, $\sim
55\%$ of the binaries from our $N=5\times10^5,1\times10^6$ models, and $\sim
35\%$ of the binaries from our $N=2\times10^5$ low-metallicity models.  Our
parameter grid contains two low-metallicity clusters for every one
high-metallicity cluster, so we further multiply the weights of the $Z =
0.05Z_{\odot},0.01Z_{\odot}$ models by 0.85 to ensure that we sample binaries
from our models assuming that 56\% of GCs are low-metallicity\footnote{Note that
this is at odds with the observed  distribution of GCs in the Milky Way, in
which $\sim75\%$ of GCs low-metallicity.  However, we find that changing the
metallicity fraction from 56\% to 75\% does not change the quantitative results
for O1 quoted in the main text, and only changes the detection rates for design
sensitivity LIGO at the $\sim 5\%$ level.} \cite[taken from the spatial
densities of low- and high-metallicity clusters determined in][Supplemental
Materials]{Rodriguez2015a}.  Finally, we note that while the integral in
equation \ref{eqn:int} is performed in redshift, we compute the KDE with the merger
times of the binaries.  This is to ensure that $P(\mathcal{M}_c,z)$ is in units of $M_{\odot}^{-1}~\rm{yr}^{-1}$.  The distribution is then expressed in redshift by assuming $P(\mathcal{M}_c,z) = P_t\left(\mathcal{M}_c,t_{\rm{lookback}}(z)\right)$, using the stated cosmological parameters.  Thus, $\mathcal{R}(\mathcal{M}_c,z)$ is in units of $\rm{Mpc}^{-3}~M_{\odot}^{-1}~\rm{yr}^{-1}$.

To compute the detectability of sources, $f_d(\mathcal{M}_c,z)$, {we assume a single Advanced LIGO detector and a minimum signal-to-noise ratio (SNR) of 8 for a detection (which is equivalent to a network SNR of 12, the current threshold for LIGO detections).  The SNR for a face-on, directly over the detector, and optimally-oriented binary is defined as:

\begin{equation}
\rm{SNR} = \sqrt{4\Re\int^{\infty}_{0}\frac{|\tilde{h}(f)|^2}{S_n(f)}df}
\end{equation}
{where $\tilde{h}(f)$ is the gravitational wave in the frequency domain and $S_n(f)$ is the one-sided power spectral density of the detector noise.  We compute $\tilde{h}(f)$ using the IMRPhenomD \citep{Husa:2015iqa, Khan:2015jqa} gravitational waveform model\footnote{This waveform model is available in the LALSimulation package of the LIGO Algorithm Library (LAL) software suite available from https://wiki.ligo.org/DASWG/LALSuite.} which covers the full BBH coalescence including the inspiral, merger, and ringdown phases.  For our effective O1 observation, we use the publicly-available power spectral density released as part of the GW150914 observation \citep{O1sensitivity-G1501223}, while our O2 and design sensitivity calculations used the noise curves cited in the main text.  These optimal SNRs (which account for the dependence on redshift and chirp mass) are then multiplied by the antenna pattern function for a single detector, $\Theta (\alpha,\delta,\iota,\psi)/4$, which describes how the gravitational wave SNR decreases for different binary sky locations, inclinations, and polarizations \citep{Finn1996}.  The actual fraction of detectable sources, $f_d(\mathcal{M}_c,z)$, is determined by computing a Monte Carlo over the domain of $\Theta$, and counting what fraction of sources at a given chirp mass and redshift have a detectable $\rm{SNR} > 8$.  See \cite{Belczynski2013,Belczynski2014} for a complete description.

What we show in Figure \ref{fig:detect} is the normalized distribution from
equation \ref{eqn:int}, where we convert from chirp mass to total mass by
assuming all sources have equal-mass components.  To compute the actual
detection rate, we integrate equation \ref{eqn:int} over all $\mathcal{M}_c$.
{Since each computation of $\mathcal{R}(\mathcal{M}_c,z)$ involves computing
$P(\mathcal{M}_c,z)$, which is based on a random draw of inspirals from our
models, we compute the integral of \ref{eqn:int} with 100 independent sample of
inspirals and report the mean of the results\footnote{This is done mostly for
completeness: our weighting scheme draws nearly all of the binaries from the
most massive clusters (which produce the most binaries), so there is
little variation between different independent samples.  In practice, the uncertainty in our estimate is primarily dominated by the uncertainty in $\rho_{GC}$.}  For our optimistic and pessimistic assumptions, we use the $\pm1\sigma$ values from the 100 independent draws, and employ the optimistic and pessimistic values of $\rho_{GC}$ and $\left<N_{\rm{insp}}\right>$ described above.  Thus, the range of values reported do not represent a true statistical uncertainty on our models, but are intended to provide a reader with a better understanding of the astrophysical uncertainties associated with our assumptions.  This procedure is similar to that developed in \cite{Rodriguez2015a}.

\end{document}